\documentclass[12pt]{iopart}

\usepackage{graphicx}
\begin{document}

\title[Density-dependent thermopower oscillations in mesoscopic 2DEGs]{Density-dependent thermopower oscillations in mesoscopic two-dimensional electron gases}

\author{Vijay Narayan$^{1,*}$, Eugene Kogan$^2$, Chris Ford$^1$, Michael Pepper$^3$, Moshe Kaveh$^2$, Jonathan Griffiths$^1$, Geb Jones$^1$, Harvey Beere$^1$ and Dave Ritchie$^1$}

\address{$^1$Cavendish Laboratory, University of Cambridge, J J Thomson Avenue, Cambridge CB3 0HE, UK}
\address{$^2$ Department of Physics, Bar-Ilan University, Ramat-Gan 52900, Israel}
\address{$^3$ Department of Electronic and Electrical Engineering, University College London, Torrington Place, London WC1E 7JE, United Kingdom}
\ead{$^*$vn237@cam.ac.uk}
\begin{abstract}
We present thermopower $S$ and resistance $R$ measurements on GaAs-based mesoscopic two-dimensional electron gases (2DEGs) as functions of the electron density $n_s$. At high $n_s$ we observe good agreement between the measured $S$ and $S_{\rm{MOTT}}$, the Mott prediction for a non-interacting metal. As $n_s$ is lowered, we observe a crossover from Mott-like behaviour to that where $S$ shows strong oscillations and even sign changes. Remarkably, there are absolutely no features in $R$ corresponding to those in $S$. In fact, $R$ is devoid of even any universal conductance fluctuations. A statistical analysis of the thermopower oscillations from two devices of dissimilar dimensions suggest a universal nature of the oscillations. We critically examine whether they can be mesoscopic fluctuations of the kind described by Lesovik and Khmelnitskii in Sov. Phys. JETP. \textbf{67}, 957 (1988).
\end{abstract}

\maketitle

\section{Introduction}
\textit{Mesoscopic} electronic systems, those with physical dimensions comparable to the Fermi wavelength $\lambda_F$, can display a variety of striking effects such as size quantisation, ballistic electron transport and quantum interference. Quantum interference effects such as Aharonov-Bohm oscillations, weak localisation and universal conductance fluctuations (UCFs) are all direct manifestations of phase coherent electron transport. Hence these are dependent on an additional lengthscale, namely the phase coherence length $l_\phi$ over which the electron loses memory of its phase through inelastic processes. UCFs are aperiodic but reproducible features in the electrical conductivity $\sigma$ as a function of an external parameter such as the magnetic field $B$ and are a consequence of the non-self-averaging nature of electron interference over lengthscales comparable to $l_\phi$. The resulting fluctuations are not limited to $\sigma$ but will tell on other transport parameters such as the thermopower as well. The thermopower or Seebeck coefficient $S$ of a system is defined as $V_{\rm{th}}/\Delta T$ where $V_{\rm{th}}$ is the voltage that develops between its end due to an imposed temperature difference $\Delta T$. $S$ and $\sigma$ are related by the Mott formula \cite{1}: 

\begin{equation}
\label{MottFormula}
S_{\rm{MOTT}} = \frac{\pi^2 k_B^2 T}{3e}\frac{\mbox{d}\ln\sigma}{\mbox{d}E}\vline_{E = \mu}
\end{equation}

Here $k_B$ is the Boltzmann constant, $T$ is the temperature, $e$ the electronic charge, $E$ the energy of the system and $\mu$ the chemical potential. Indeed, ‘universal thermopower fluctuations’ have been experimentally demonstrated in quantum wires \cite{2}, quantum dots \cite{3} and, more recently, in other mesoscopic structures \cite{4}. While mesoscopic fluctuations have been reported most often as a function of $B$, they are expected \cite{4, 5, 6} and seen \cite{7, 8, 9} even as a function of $\mu$. The relative magnitude of the $S$-fluctuations can be much larger than those in $\sigma$ by virtue of the derivate relation in equation \ref{MottFormula}. Lesovik and Khmelnitskii (henceforth LK) show \cite{10} that precisely for this reason, the mesoscopic contribution to the thermopower can be exceedingly large and either positive or negative. Consequently, they can potentially swamp the regular, ‘bulk’ contribution even causing sign reversals in $S$. The important difference between such fluctuations and thermopower sign-reversals resulting from Coulomb Blockade \cite{11, 12, 13} is that the former occur in the high-conductivity regime $\sigma >> h/e^2$, while the latter become significant at $\sigma \sim h/e^2$.

Recently some of us reported $S$ measurements in low-density two-dimensional electron gases (2DEGs) of mesoscopic dimensions \cite{14, 15} as a function of density $n_s$ and temperature $T$. Specifically, $S$ displayed large oscillations and even sign changes as the 2DEG density $n_s$ was varied whereas the electrical resistivity $\rho \equiv 1/\sigma$ was completely monotonic. However, these measurements focused on the low-conductivity regime of the samples (300$h/e^2 > \rho > 5h/e^2$), where many-body effects suppressed the strong localisation (see also refs~\cite{16} and \cite{17}) and it is not clear whether weak localisation, and therefore, mesoscopic fluctuations are to be expected. In this work we measure $S$ in similar samples while concentrating specifically on the medium-$\rho$ range around the onset of $S$-oscillations where, in principle, UCFs are expected. We perform a statistical analysis of these oscillations and examine whether they are consistent with the mesoscopic fluctuations described by LK.

The remainder of the article is structured as follows: Section 2 describes the experimental system and measurement setup, section 3 presents the experimental data and statistical analysis and Section 4 presents a discussion and concluding remarks.

\section{EXPERIMENTAL DETAILS}

\begin{figure}
	\centering
	\includegraphics[width=4in]{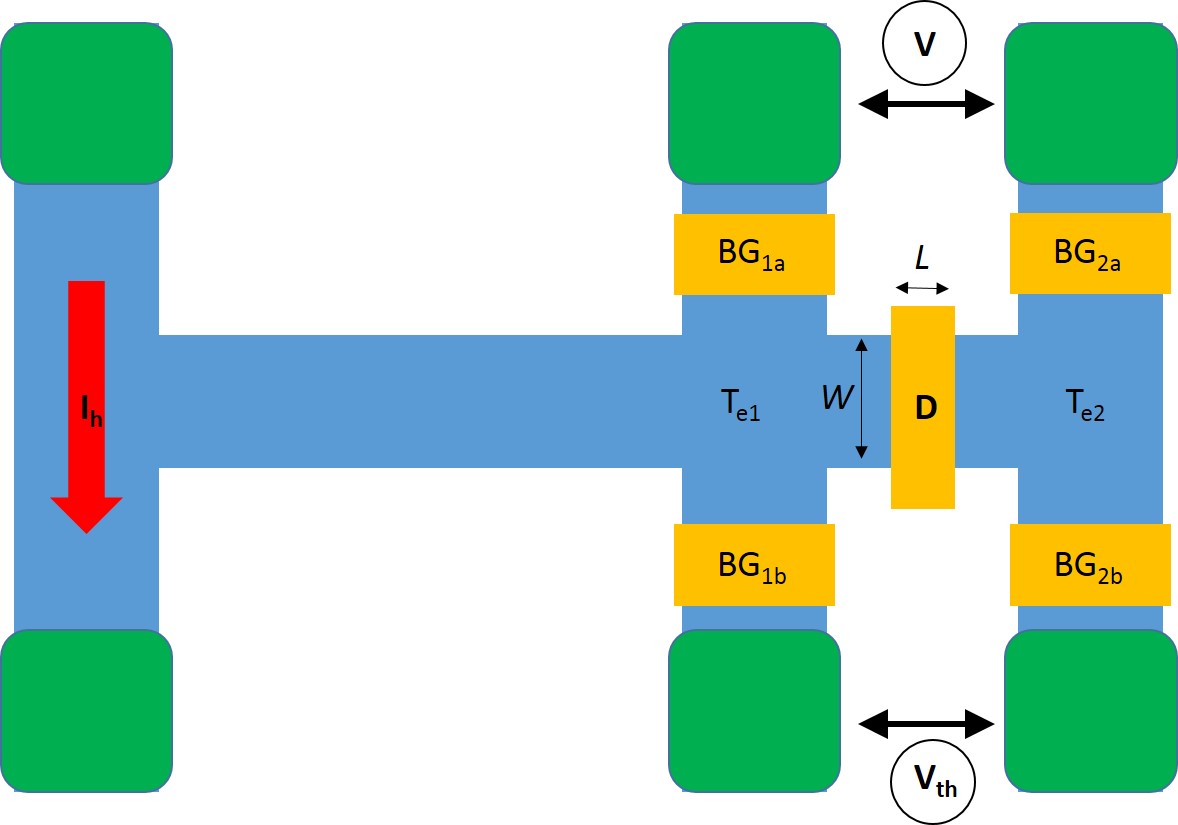}
	\caption{Schematic representation of the device and measurement scheme. The blue area represents the conducting mesa and ohmic contacts are represented in green. The yellow rectangles represent top-gates.}
	\label{fig1}
\end{figure}

Our experiments are all performed at $T$ = 0.3~K in GaAs-based 2DEGs with an as-grown mobility of 220~m$^2$/Vs at $n_s$ = 2.1$\times$10$^{15}$~m$^{-2}$. We use a wet etch to define a conducting mesa and deposit ohmic contacts to perform electrical and thermoelectric measurements. The mesoscopic device D (see figure 1) is defined by a metallic gate of lithographical dimensions L$\times$W = $2 - 3~\mu \rm{m} \times 8~\mu \rm{m}$. Figure \ref{fig1} shows a schematic representation of the device. In addition to D, there are four gate-defined bar-gates (BGs) that serve as a pair of thermometers to measure the local electron temperature $T_{\rm{e}}$on either side of D~\cite{14}. To measure $S$ we heat one end of the device with an AC heating current $I_{\rm{h}}$ = 4~$\mu$A at frequency $f_{\rm{h}}$ = 11~Hz. $V_{\rm{th}}$ is measured using a lock-in amplifier at 2$f_{\rm{h}}$ as shown in figure~\ref{fig1}. The local electron temperatures on either side of the device $T_{\rm{e1}}$ and $T_{\rm{e2}}$ are measured using bar-gate pairs BG$_{\rm{1a,b}}$ and BG$_{\rm{2a,b}}$, respectively. Details of this measurement procedure can be found in Ref.~\cite{14}. $S$ is then calculated as $V_{\rm{th}}/(T_{\rm{e1}} - T_{\rm{e2}})$. We measure the resistance $R$ using a standard four-probe lock-in technique with a constant current $I_{\rm{ex}}$ = 1~nA at a frequency $f$ = 7~Hz. However, the results are unaffected upon increasing $I_{\rm{ex}}$ by an order of magnitude to 10~nA. In all our measurements $V_{\rm{th}}$ and $R$ are measured simultaneously. $\Delta T$ is 20~mK in all the measurements reported here.

\section{Experimental Data}

\begin{figure}
	\centering
	\includegraphics[width=6in]{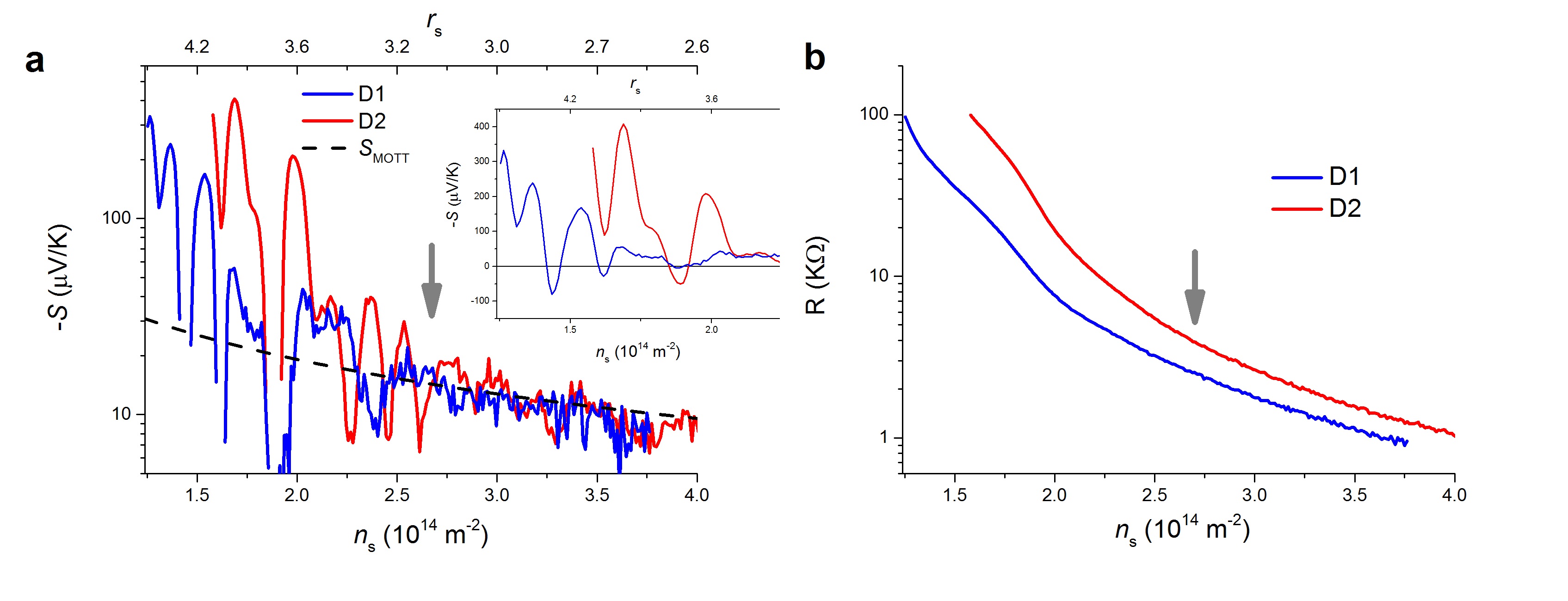}
	\caption{(a) $S$ as a function of $n_s$ (bottom axis) and $r_s$ (top axis). The inset shows the same data on a linear scale where the sign-reversals of $S$ are clearly visible. Figure 2b shows $R$ measured simultaneously with $S$ over the same $n_s$. Blue and red traces correspond to devices with L = 2~$\mu$m and 3~$\mu$m, respectively and the arrow marks the density below which thermopower oscillations are present.}
	\label{fig2}
\end{figure}

Figures \ref{fig2}a and \ref{fig2}b show $S$ and $R$ as functions of $n_s$ for two different samples labelled D1 and D2. The top horizontal axis in figure \ref{fig2}a shows the interaction parameter $r_s \equiv 1/a_B\sqrt{\pi n_s}$, where $a_B$ is the effective Bohr radius in GaAs $\approx$ 11~nm. D1 has L = 2~$\mu$m and D2 has L = 3~$\mu$m. Figure \ref{fig2}a also shows $S_{\rm{MOTT}}$ for $\sigma = n_se^2\tau/m$

\begin{equation}
\label{DrudeMott}
S_{\rm{MOTT}} = \frac{\pi^2 k_B^2 T}{3e}\frac{\mbox{d}\ln\sigma}{\mbox{d}E}\vline_{E = \mu} = \frac{\pi k_B^2 T m}{3e \hbar^2} \frac{1 + \alpha}{n_s}
\end{equation}

\begin{figure}
	\centering
	\includegraphics[width=6in]{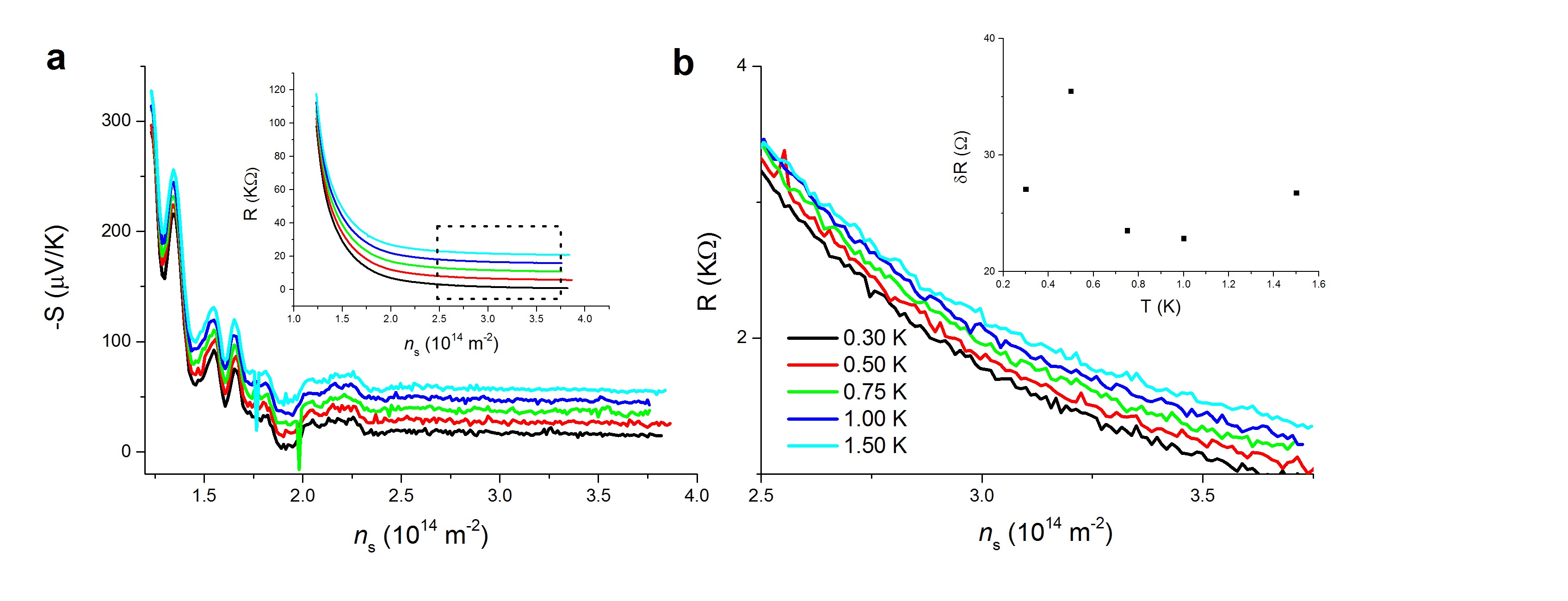}
	\caption{(a) shows five separate $S$ traces vertically offset to each other. The inset shows the corresponding $R$ traces. All data are taken at $T$ = 300~mK. Figure 3(b) shows five $R(n_s)$ traces, again, with a relative vertical offset, at five different $T$ and concentrating on high-$n_s$ (boxed region in the inset to figure 3(a)). The inset shows the scale of fluctuations $\delta_R$ as a function of $T$.}
	\label{fig3}
\end{figure}

Here $\tau$ is the inelastic scattering time and $\alpha = (n_s/\tau)(\mbox{d}\tau/\mbox{d}n_s) \approx$~1. It is seen that $S$, while agreeing closely with $S_{\rm{MOTT}}$at high $n_s$, shows a sudden departure from Mott-like behaviour at $n_s \approx$~2.5$\times$10$^{14}$~m$^{-2}$ with strong oscillations and even sign reversals. The sign reversals are seen most clearly in the inset to figure \ref{fig2}a which also serves to distinguish the smooth $S$ oscillations from the measurement noise ($\sim$10~$\mu$V/K which corresponds to a voltage uncertainty $V_{err}$ of 200~nV), the latter being most prominent in the range $n_s >$ 2$\times$10$^{14}$~m$^{-2}$. It is worth emphasising that the former are large oscillations and completely reproducible between $n_s$ sweeps (see figure \ref{fig3}), while the latter correspond to measurement noise. In the same $n_s$-range, $R$ grows monotonically in both devices except for some broad features accentuated by the log scale which, importantly, bear no obvious correlation to the oscillations in $S$. Hence, this very unusual behaviour cannot simply be reconciled with $S$ oscillations in a disordered 2DEG that arise, say, due to Coulomb blockade resulting from the 2DEG fragmenting into charge puddles.

In figure~\ref{fig3}a we plot five consecutive $S$ traces taken at $T$ = 0.3~K vertically offset relative to each other. In the inset we plot the simultaneously measured $R(n_s)$ corresponding to each $S(n_s)$ trace, again, with a relative vertical offset. These figures firmly establish both the reproducible nature of the $S$ oscillations as well as the smooth character of $R(n_s)$. In figure~\ref{fig3}b we investigate whether the $R$ data at high-$n_s$ contain any signatures of UCFs. We see that in the five $R(n_s)$ traces at 0.3~K $< T <$ 1.5~K, vertically offset relative to each other by 100~$\Omega$, there are no reproducible features but a fluctuation level that has no obvious $T$-dependence. This becomes clearer in the inset where we plot $\delta R$, the magnitude of fluctuations in $R$ as a function of $T$. $\delta R$ is defined as the standard deviation in $R$ after subtracting a smooth background. The dependence of $\delta R$ on $T$ is not consistent with weak localisation and $\delta R$ simply represents the experimental noise levels.

\begin{figure}
	\centering
	\includegraphics[width=6in]{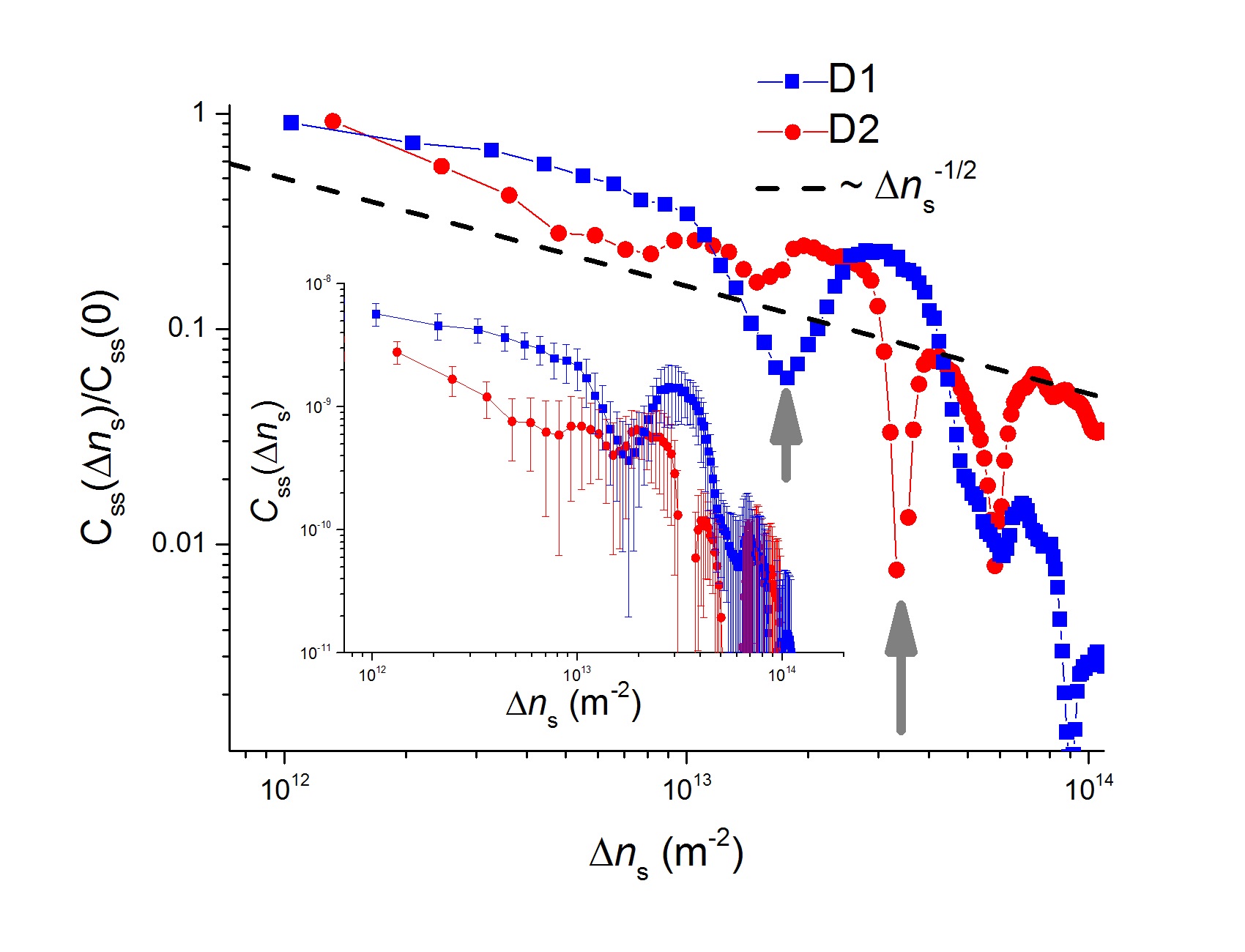}
	\caption{The correlation function $C_{SS}(\Delta n_s)$. The grey arrows indicate the minimum in each instance. The inset shows the same data without normalisation.}
	\label{fig4}
\end{figure}

\begin{figure}
	\centering
	\includegraphics[width=6in]{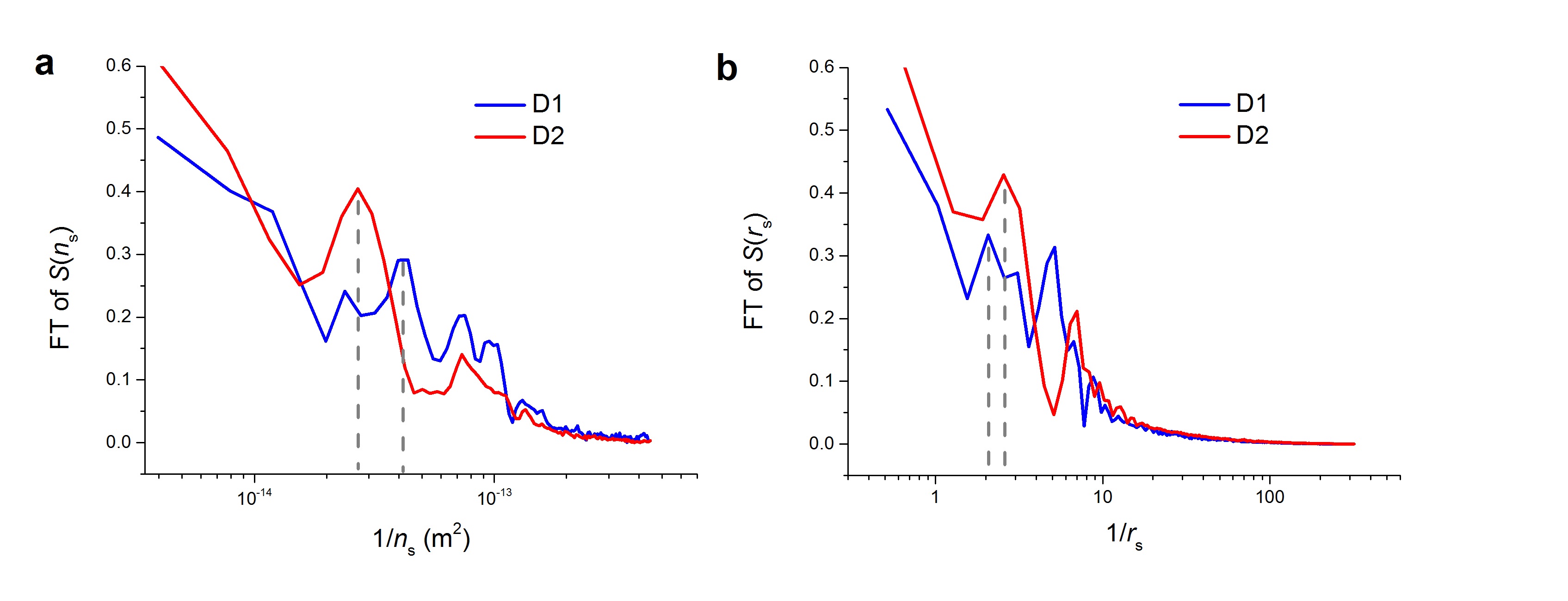}
	\caption{Figures 5a and 5b show the fourier transform of the data in figure 2(a).}
	\label{fig5}
\end{figure}

We now return to the $S$ oscillations and to analyse their statistical nature, we look at the autocorrelation of $S$. In figure~\ref{fig4} we plot $C_{SS}(\Delta n_s) \equiv \langle\delta S(n_s) \delta S(n_s + \Delta n_s)\rangle/R^2(n_s)$ normalised to $C_{SS}(0)$ = 1. Here the angular brackets represent an average over $n_s$ and $\delta S(\Delta n_s)$ is defined as $S$ - $S_{\rm{MOTT}}$.  The correlator is defined with $R^2(n_s)$ in the denominator rather than $S^2(n_s$) to avoid divergences at the zero-crossings of $S(n_s)$ and also to make a quantitative comparison with the LK theory (see next section). The inset shows the data without normalising and error bars as estimated from the standard deviation in the data. We find that $C_{SS}(\Delta n_s)$ in both devices agree quantitatively and are consistent with an initial decay $C_{SS}(\Delta n_s) \sim \Delta n_s^{-1/2}$. Furthermore, there is a minimum and subsequent maximum that occur in the vicinity of $\Delta n_s \approx$ 2.5$\times$10$^{13}$~m$^{-2}$. These features suggest a degree of periodicity in the $S$ oscillations which are further reflected in the fourier transforms (FTs) of the data shown in figure~\ref{fig5}. The grey arrows mark the FT peaks which correspond to periodicities of $\Delta n_s \approx$~4.2$\times$10$^{-14}$~m$^{-2}$ and 2.7$\times$10$^{-14}$~m$^{-2}$ for D1 and D2, respectively. We note that figure~\ref{fig4} shows several features at $\Delta n_s$-values larger than 5$\times$10$^{-13}$~m$^{-2}$ which can be correlated to peaks in the FT in figure~\ref{fig5}a. However, these must be taken with caution due to the lack of sufficient statistics. In figure~\ref{fig5}b we plot the FT of $S$ as a function of $r_s$. Interestingly we note that the periodicities correspond to $\Delta r_s \approx$ 0.4 –- 0.5, reminiscent of conductance oscillations seen in disordered silicon inversion layers~\cite{18}.

\section{Discussion}

The strong oscillations in $S$ without corresponding features in $R$ is extremely surprising and signifies a breakdown of the Mott formula (equation \ref{MottFormula}). While there are instances where this is expected (see section 5), we first ascertain whether the mesoscopic nature of the 2DEG has a role to play in the observed experimental data. As mentioned earlier, mesoscopic fluctuations are necessarily present in all transport parameters of mesoscopic samples. However, oscillations in $S$ can be significantly amplified compared to those in $R$ due to the derivative relation between them. Conversely, therefore, one can expect situations where $S$ oscillations are visible but immeasurably small in $R$ and such situations would be wrongly perceived as a breakdown of the Mott relation. Motivated by this we now investigate whether the observed oscillations are consistent with the LK-type oscillations~\cite{10} mentioned earlier.

LK calculated the mesoscopic contribution to $S$ of a sample of small size as a function of $\mu$ and the magnetic field $B$. Using the Mott formula (equation~\ref{MottFormula}) they then showed that this contribution can be  positive or negative and even exceed the regular term in magnitude, potentially resulting in sign reversals of $S$. Furthermore, they constructed and evaluated the correlation function: $C_{SS}^{LK}(\Delta \mu, \Delta B) \equiv \langle\delta S(0,0) \delta S(\Delta \mu, \Delta B)\rangle = (R^2/\hbar^2)F_S(\Delta \mu \tau_F/\hbar, \Delta B LW/\phi_0)$. Here $F_S$ is a scaling function, $\tau_F$ is the time taken for an electron to diffuse across the sample of length $L$ and is given by $\tau_F \approx L^2/D$, with $D$ being the electron diffusivity and $\phi_0 = hc/e$. At $\Delta B$ = 0 and low $T$ this correlation functions is essentially identical to the correlation function evaluated in figure~\ref{fig4}. The low-$T$ requirement is in order to approximate $\mu$ by the Fermi energy $E_F$ and this is met in our experimental results since $k_BT/E_F \leq$ 0.25 in the $n_s$-range being considered. LK predict a universal form for the autocorrelation function which they numerically show to initially decay and then go through a minimum. This is in striking agreement with the result in figure~\ref{fig4} in that the data from both devices has the same initial $\Delta n_s^{-1/2}$ decay and a clear minimum.

Additionally, LK consider the cross-correlation between $S$ and $R$: $C_{SR}^{LK}(\Delta \mu, \Delta B) \equiv \langle\delta S(0,0) \delta R(\Delta \mu, \Delta B)\rangle$, where $\delta R$ are fluctuations about the mean $R$. LK show $C_{SR}^{LK}(\Delta \mu, \Delta B)$ to be 0 at $\Delta \mu$ = 0, i.e., oscillations in $S$ and $R$ at the same value of chemical potential are uncorrelated, but gradually grow as a function of $\Delta \mu$ and ultimately go through a maximum. While the statement that $S$ and $R$ oscillations at a given $\mu$ are uncorrelated seems consistent with a visual inspection of the data in figure~\ref{fig2}, a quantitative comparison reveals discrepancies with the LK picture. First, at the point where $S$ oscillations set in, there are absolutely no oscillations in $R$ (see figures~\ref{fig2} and \ref{fig3}), thereby trivially rendering $C_{SR}$ zero. In this region, $R \approx$ 3~K$\Omega$ and therefore the expected magnitude of $\delta R = e^2R^2/h \approx$ 350~$\Omega$. This is well within the limit of accuracy even if, considering the wide aspect ratio of the sample, we divide the expected $\delta R$ by a factor $\sqrt{W/L} \approx$~2. Given the above estimate for $\delta R$, it is then reasonable to ask why UCFs are absent in the first place. While this is unclear at the moment, a possible explanation lies in the nature of the disorder being sampled by the device. The earlier experiments on mesoscopic 2DEGs~\cite{16} that motivated the present studies were performed especially to circumvent the long-range disorder due to remote ionised dopants in the host GaAs wafer. The residual short-scale ‘white’ disorder would then restore the self-averaging nature of the electron paths, leading to the absence of UCFs. We hope to verify this argument by examining whether UCFs appear in larger mesoscopic samples.

It is worth mentioning here that preliminary results of $S$ and $R$ as a function of $B$ also indicate the absence of UCFs. Not only is this consistent with with the arguments in the previous paragraph, it is another discrepancy between the experimental data and the LK picture. The LK analysis predicts similar cross-correlation and autocorrelation functions as $B$ is tuned, but we observe no reproducible oscillations in $S$ or $R$ as functions of $B$, or at least no oscillations (other than the $B^{-1}$-periodic Shubnikov-de Haas oscillations) comparable in magnitude to those observed by tuning $n_s$. This data will be presented in a separate report.

\section{Conclusions}

To summarise, we find some aspects of the data to be consistent with oscillations due to the mesoscopic nature of the device, but several others that are at odds with it. Furthermore, the $S$ oscillations persist to much lower $n_s$ even up to $R \approx$~100~$h/e^2$~\cite{14,15}, a regime in which there is no reason to expect LK-type oscillations. While the oscillations may indeed be related to those seen in Ref.~\cite{18}, there have recently been several theoretical reports describing situations where the Mott relation (equation~\ref{MottFormula}) breaks down. These include the vicinity of a quantum critical point~\cite{19}, proximity to a Lifshitz transition~\cite{20}, and, remarkably, even far away from a critical point~\cite{21}.

In conclusion, we have observed intriguing oscillations in the thermopower $S$ of mesoscopic 2DEGs and analysed their statistical nature. We find that the experimental data from the 2 devices measured suggests several common aspects in the nature of oscillations despite the devices being of dissimilar sizes: 1. the decay of the autocorrelation function of $S$ is consistent with a $\Delta n_s^{-1/2}$ dependence; 2. the minima and maxima of the autocorrelation in the two devices occur at approximately similar locations; and 3. the oscillations have a degree of periodicity as revealed by the fourier transform. We have compared these results to the oscillations predicted by Lesovik and Khmelnitskii in Ref.~\cite{10} and though there are suggestive similarities between the two, it seems unlikely that they are related.

\section{Acknowledgements}

We acknowledge funding from the Leverhulme Trust and EPSRC, UK. VN acknowledges a Fellowship from the Herchel Smith Fund, University of Cambridge. VN also acknowledges useful discussions with Sriram Shastry and Charles Smith.

\end{document}